\documentclass[12pt,preprint]{aastex}

\shorttitle{Spatially Resolved Nuclear Dust in Centaurus A}
\shortauthors{M. Karovska et al.}

\begin{document}


\title{Spatially Resolved Circumnuclear Dust in Centaurus A}

\author{Margarita Karovska, Massimo Marengo, Martin Elvis, Giovanni
G. Fazio, Joseph L. Hora}
\affil{Harvard-Smithsonian Center for Astrophysics, 60 Garden St.,
Cambridge, MA 02138}
\email{karovska@cfa.harvard.edu}

\author{Philip M. Hinz, William F. Hoffmann, Michael Meyer, Eric Mamajek}
\affil{Steward Observatory, University of Arizona, 933 North Cherry
Avenue, Tucson, AZ 85721-0065}


\begin{abstract}
In this paper we present results from our exploratory mid-IR study of
Centaurus A circumnuclear environment using high-angular resolution
imaging at the Magellan 6.5m telescope with the MIRAC/BLINC camera. We
detected emission from a compact region surrounding the nuclear
source, and obtained photometry at 8.8~$\mu$m and in the N band. Our
analysis suggests that the nuclear region is resolved with a size of
$\approx$3~pc. The mid-IR emission from this region is likely
associated with cool dust with an estimated temperature of $\sim$160~K,
surrounding the central ``hidden'' AGN. We discuss the characteristics
of this emission in relation to other mid-IR observations and 
the implications on models of dust formation in AGNs.
\end{abstract}




\section{Introduction}\label{sec-intro}

The nearest radio-bright AGN is located at a distance of ``only''
$\sim 3.5$~Mpc in the giant elliptical galaxy
NGC5128 (Cen A) ({\it vis}
review by Israel 1998). The central
engine powering this galaxy is thought to be a supermassive
black hole 
enshrouded in a region containing dust and gas 
heated by the central source (Marconi {\it et al.}
2001).
Multi-wavelength observations of Cen A
reveal many complex
multi-scale structures, at scales ranging from a sub-pc to hundreds of
kpc (eg. Karovska {\it et al.} 2002).
A radio and
X-ray jet powered by the central source  extends across
the galaxy to a distance of
several hundreds of kpc
from the nucleus. 
Optical and near-IR images of Cen A show large spatial scale (several
kpc in size) dark bands stretching across the
middle of the
galaxy and obscuring the central
region, probably due to
absorption by dust and other cool material. These dust lanes are
 thought to be a remnant of a merger with a smaller spiral
galaxy (Schiminovich {\it et al.} 1994).
Using mid-infrared and
sub-millimeter wavelength observations of the  central region of Cen A,
Mirabel {\it et al.} (1999) detected 
dust emission from a large scale
bisymmetric structure (5kpc in size) resembling a 
barred spiral galaxy.

Because of its proximity, Cen A provides a unique possibility for
high-spatial resolution 
studies of the circumnuclear
environment of AGNs using ground- and space-based large-aperture
telescopes (0.1'' corresponds to $\sim 1.5$~pc at the Cen A distance
of $\sim$3.5~Mpc).
However, exploring the nuclear region of Cen A at optical
wavelengths presents a major challenge because of the high extinction
(eg. $A_V \sim 10$~mags; Alexander et al. 1999) due to
heavy  obscuration by the dust in the dust lanes.

Although the diffraction limited telescope resolution is significantly
lower in the mid-IR when compared to the optical ($\sim$ 10 times), 
the spectral range from 8 to 20~$\mu$m is, nevertheless,
particularly suitable for studying the emission from the region
surrounding the nucleus because of at least ten times lower opacity.
The lower opacity of astronomical dust at these wavelengths
(Draine \& Lee 1984) permits penetration of
the optically opaque dust lanes.
Most of the
thermal radiation from dust is in fact emitted at these wavelengths,
as seen in the mid-IR ISOCAM spectra of Cen A which 
show the presence of the characteristic silicate broad absorption
feature at 9.8~$\mu$m, plus a possible PAH emission line at
11.3~$\mu$m. 
Alexander et al. 1999 suggested that the presence of a very 
deep 10~$\mu$m silicate absorption
feature in the ISOCAM spectrum could be due to self-absorption in a
dusty circumnuclear region, in addition to the extinction by the dust in the
dust lanes (optical depth $\tau_\nu \simeq 1$ at 10~$\mu$m).

We report here results from our mid-IR
imaging of Cen A with the MIRAC/BLINC mid-IR camera at
Magellan, showing that the nuclear region is resolved at a
sub-arcsecond scale.
The
observations and the techniques for data acquisition and reduction are
described in the next section. Results and analysis are presented in
section 3, and discussed in relation to other
available observations and models in section 4.


\section{Observations and Data Reduction}\label{sec-obs}

During an exploratory observation of Cen A on May 1 and 2, 2002 at the
Magellan I (Baade) telescope we detected a 
compact mid-IR source 
in the images of the nuclear region at 8.8~$\mu$m and in the N
band. The Baade Telescope, with a primary aperture of 6.5m equipped
with active optics, provides diffraction limited sub-arcsecond images in the
8.8~$\mu$m spectral window (Shectman \& Johns 2003).
The images of Cen A and several reference sources were recorded using 
the mid-IR camera MIRAC/BLINC. MIRAC/BLINC uses a Boeing HF16 128$\times$128 Si:As
blocked impurity band detector (Hoffmann et al. 1998). 

We obtained images
using the MIRAC/BLINC 8.8~$\mu$m 10\%
passband filter and the wide N band filter (from 8.1 to 13.1~$\mu$m)
centered at 10.3~$\mu$m. The 
total on-source integration time was 1400~s at 8.8~$\mu$m and  900~s 
in the N band. The reference stars
$\gamma$~Cru and IRC$+$10220, observed while transiting at a similar
airmass as the source, provide flux and Point Spread Function (PSF)
calibration information. 
To insure similar observing conditions, including similar
telescope orientation, airmass and seeing conditions, 
the observations of Cen A were made 
close in time (within couple of hours) to the observations of the corresponding
reference stars. 
In addition, we used the Magellan telescope
active optics (looking at standard stars in
the field of view) to monitor the seeing stability, by checking that the 
size and the 
shape of the PSF at the beginning of each observation of Cen A is consistent
with those of the corresponding reference. 

On the Magellan,
MIRAC/BLINC has a plate scale of 0.12~arcsec pix$^{-1}$, providing a total
field of view of $17'' \times 17''$. This pixel scale ensures Nyquist
sampling of the diffraction-limited PSF. 
As described below, we obtained further 
sub-sampling of the PSF by using dithering techniques.
We used a standard nodding and chopping technique to remove the
background signal, dithering the source on the array to obtain 
sub-sampling of the PSF. 
Chopping is carried
out through rotation of an internal mirror to mitigate noise
due to sky brightness fluctuations.
The chop frequency was set to 10~Hz, with a
throw of $8''$ in the North-South direction. The nod throw was also set
to $8''$, but in the East-West direction, in order to have all four
chop-nod beams inside the field of view of the array. Each individual
nod cycle required 15~s (8.8~$\mu$m) or 10~s (N band) on-source
integration, and the procedure was repeated for as many cycles as
needed to obtain the total integration time at each wavelength. 

The data were analyzed by first subtracting the chop-on from the
chop-off frames for both nodding beams. The two images thus obtained
were then subtracted one from the other, in order to get a single
frame in which the source appears in all four beams (two negative and
two positive). We then applied a gain matrix, derived from images of
the dome (high intensity uniform background) and the sky (low
intensity uniform background), to flat field the chop-nodded image. 
This procedure was repeated for each of the nodding cycles for which
the source was observed. A final high signal-to-noise ratio cumulative
image was then obtained by co-adding together all the beam frames, each
re-centered and shifted on the source centroid. This last 
co-adding step was performed on a sub-pixel grid, providing a final pixel
scale of 0.03~arcsec pix$^{-1}$. A mask file to
block out the effects of bad pixels and field vignetting was also
created and applied at each individual frame before their shifting,
preventing individual rejected
pixels from contributing to the final image. 
To ensure a
uniform treatment of the source and the standard the same observing and
reduction procedure was also used for the reference star.


\section{Results and Analysis}\label{sec-results}

\subsection{Imaging of the Circumnuclear region}

The mid-IR images of the nuclear region of Cen A obtained at
8.8~$\mu$m and in the N band are 
systematically larger then the unresolved reference
sources (or the ``observed'' PSFs) recorded under similar observing 
conditions and using the same filters.
In Figure~1 we show the 
images of Cen A and the reference star $\gamma$~Cru obtained at 8.8~$\mu$m.
The comparison of image sizes clearly shows that the Cen A image
of the nuclear region is 
larger than the image of the unresolved point-like reference source 
at this wavelength.
We note that the 8.8~$\mu$m images of Cen A and the reference
 have better resolution
and are of higher signal-to-noise then the images obtained in the wide N band
centered at 10~$\mu$m.

The difference in size between the point source and the Cen A nuclear
region is shown in Figure~2, which displays the radial profiles of the
source and the reference at 8.8 ~$\mu$m normalized at the same peak value.
This suggests that the Cen A nuclear region
is resolved, with the size near
the resolution limit of the Magellan 6.5 m telescope.
We estimated the angular size of the Cen A nuclear region by 
convolving the reference star radial profile with a gaussian, and then
fitting to the Cen A data at the corresponding bandpasses.
The best fit at 8.8~$\mu$m gives a FWHM of 0.17 $\pm 0.02''$.
The N band size estimate is consistent, with larger error bars
because of the lower S/N in these data.

At the distance of Cen A, the measured FWHM angular size
corresponds to a 
linear size of $\approx 3$~pc, or $\sim$10~Ly.
This
is in agreement with the predicted size of the dust emitting region
based on model fitting of ISOCAM spectra with 
emission from the dust lanes and from a separate toroidal
dust region around the nucleus with a diameter of $\sim 3.6$~pc
(Alexander et al. 1999).

\subsection{Photometry}

We obtained photometry of the nuclear region
 using Cen A images in the two observed
bandpasses by selecting a small aperture centered on the
nuclear region. We used an aperture with a diameter of 20
MIRAC/BLINC pixels, corresponding to $\sim 2.4$~arcsec, which is the
minimum aperture size which includes most of the PSF flux. 
The photometric reference for the 8.8~$\mu$m image was 
$\gamma$~Crux, which has a known flux of 1090~Jy
at that wavelength, with an estimated uncertainty of 5\%
(Gezari {\it et al.} 1987). For the N band we used the source IRC$+$10220 as a
reference. We estimated for this star an N band magnitude of $\sim
1.5$~Jy, with a 30\% uncertainty, based on interpolation of the K magnitude
of 2.84 ($\sim 45$~Jy) and the IRAS 12~$\mu$m flux of $\sim 4$~Jy from
Gezari {\it et al.} (1987) (Table~1).

The photometry of the Cen A nuclear region was calculated separately
for the four nod/chop beams at each wavelength, which were then 
averaged together. This procedure
ensured that the variation in the photometry of the four beams was
within our error estimate. 
We also evaluated different sizes of the aperture and of the sky region
for residual background subtraction, from 1.5~arcsec to 5~arcsec,
finding similar values of the photometry, within our error bars ($\pm$1$\sigma$).
We obtained an average flux of 0.9~Jy at
8.8~$\mu$m, with an uncertainty of $\pm 0.1$~Jy, and a flux of 1.4~Jy
in the N band, with an uncertainty of $\pm 0.5$~Jy (Table~1).
The significant uncertainties are due to the limited photometric
quality of the sky during these observations.
Larger aperture photometry 
(several arcseconds aperture) does
not show significant differences within the error bars
with the smaller aperture results that
we are presenting in this paper.

Our photometric results are in agreement 
with the mid-IR observations obtained by
 Whysong \& Antonucci (2002) using the Keck I telescope.
They detected emission from a compact unresolved nuclear source
(with an upper limit on the
size of $\sim 0.3''$) and measured fluxes of 1.6~Jy and 2.3~Jy  
at 11.7~$\mu$m and 17.75~$\mu$m,  respectively.

Our results are also in agreement within $\approx 3\sigma$
errors
with the ISOCAM CVF
spectrum obtained by Mirabel et al. (1999), which gives $\sim 0.7$~Jy
at 8.8~$\mu$m. In Figure 3 we plot our photometric results on
the ISOCAM CVF spectrum (see Fig. 2 in Mirabel et al. 1999). 
The spectrum shows a broad
silicate absorption feature around 9.8~$\mu$m, resulting in a
lower flux at 10.3~$\mu$m with respect to 8.8~$\mu$m. 

Given the limited photometric and spectral accuracy of our mid-IR data
we cannot
discriminate between thermal emission from dust 
and the emission from late-type stars photospheres, because we
essentially have only one color (the 8.8 micron filter
is inside the N filter waveband). It is also likely that
the eventual stellar radiation in the nuclear region is
below our sensitivity limit.
We therefore assume that the emission in the circumnuclear
region is thermal, originating from dust surrounding the inner hot region.

\subsection{Dust Temperature}

Using the estimated size of the
circumnuclear region at 8.8$\mu$m and the flux measurements
based on our mid-IR observations, we derive an estimate
of the temperature of the emitting dust.
We assume that the emission is a  black body
thermal emission, and that there is an absorbing component associated with the
dust lanes in front of the central source. 
For
a simple spherical
geometry of the emitting region 
we
estimate the dust temperature at 8.8~$\mu$m using the following
expression
derived from the Planck law:

\begin{equation}
T= 1440 \ \left[ \ln \left( 1+ 1.1 \cdot 10^5 \ \frac{R^2  \
   e^{-\tau_\nu}}{D^2 \ F_\nu} \right) \right]^{-1} 
\end{equation}

\noindent
where $T$ is the dust temperature in K, $R$ is the radius of the region in
pc, $F_\nu$ is the
flux in Jy, and $\tau_\nu$ is
the optical depth of the dust lanes.

Using a radius of $\approx 1.5$~pc for the circumnuclear dust region,
flux of $0.9$~Jy, and $\tau_\nu \sim 1$ for the dust lanes optical
depth at $\sim 10$~$\mu$m,  we estimate the dust temperature at
$\approx 160$~K.

We compare the temperature obtained using this simplified model with
the temperature derived from the ISOCAM spectra.
The multicomponent spectral fit (see Figure~3  in Alexander et al. 1999)
shows that the mid-IR emission from the circumnuclear region peaks at
20~$\mu$m, which corresponds to a dust temperature of $T \simeq 150$~K.
This is in agreement with the estimated dust temperature of the
circumnuclear region based on our mid-IR
measurements.

\
\section{Conclusions}

The results of our mid-IR imaging and photometry of the Cen A nuclear
region suggest that the emission originates from a resolved source
$\approx$3~pc in diameter. This emission is likely to be thermal
radiation from dust at $\approx$150~K
heated by  UV/optical radiation from a ``hidden'' central AGN source
(e.g. Peterson 1997; Whysong \& Antonucci 2002). It is
possible that we
are detecting the external cooler region, possibly surrounding hotter
inner dust ($T \sim 1000$~K), as suggested by the deep silicate
absorption feature in the ISO spectra around 10~$\mu$m.

Resolving the source of mid-IR emission in the nuclear region of Cen A
provides an important key for understanding dust formation
processes  and the structure of the nuclear region of AGNs and quasars.
For example, the current ``unified'' model of AGNs predicts a 
dusty torus with a size of $\sim 10$~Ly surrounding the accretion disk
and a black hole central engine (Antonucci and Miller 1985; Krolik
and Begelman 1988; Peterson 1997). The predicted size of the torus  is
of the order of the measured size of the Cen A circumnuclear region
using our `mid-IR imaging. We note that the estimated size of the Cen A
circumnuclear region  is also similar to the size of the dusty region
($\sim 3$~pc) surrounding the central source of NGC~1068, recently
resolved AGN using the VLTI observations (ESA Press release 17/03).

The measured size of the mid-IR emitting region is also 
comparable to the distance from the central source at
which dust will condense in an outflowing
AGN wind.  As recently demonstrated by Elvis {\it et al.} (2002), 
dust can be created from cooling Broad Emission Line
clouds in AGN winds, and can survive in the quasar
environment due to self shielding. 
This implies that Cen A could effectively be recycling dust from the 
host galaxy
interstellar medium into IGM while changing the dust size and
distribution and the dust-to-gas ratio.
The possibility that AGN
or quasars
can provide 
an additional path
for dust formation has important cosmological implications, since it 
could explain in a
natural way the observed heavy obscuration of distant quasars (Omont
et al. 2001; Elvis {\it et al.} 2002),  and provide a new means of
forming dust at early cosmological times. 

The current observations do not have enough accuracy to determine the
true geometry of the emitting region or to distinguish between
different geometries suggested by AGN models.
The observations were challenging because of the
limited telescope resolution.
Further multi-wavelength observations of Cen A with higher spatial
resolution
are crucial to explore the geometry of the extended nuclear region and
derive the
physical characteristics and the origin of the emission.
Multi band mid-IR observations combined with spectroscopy will allow
determination of the
characteristics of the circumnuclear dust, including its temperature,
and an estimate of the mass and mass loss rate.
The results may further constrain the unified AGN model, and models of dust
formation in quasar outflows.


\acknowledgements
We thank the Magellan staff for their outstanding support. We are
grateful to
Dr. Scott Wolk for valuable suggestions and aid in preparing the
manuscript, and to the anonymous referee for useful comments
and suggestions.
M.~K. and M.~E. are
members of the Chandra X-ray Center, which is operated by the
Smithsonian Astrophysical Observatory under contract to NASA
NAS8-39073. 
This research was supported in part by NASA through the American
Astronomical Society's Small Research Grant Program.
MIRAC is supported through SAO and NSF grant AST 96-18850. BLINC is
supported through the NASA Navigator program.

\clearpage


\clearpage


\begin{table}
\begin{scriptsize}
\begin{center}
\begin{tabular}{lccc}
\multicolumn{3}{c}{\scriptsize TABLE 1}\\
\multicolumn{3}{c}{\scriptsize MID-IR PHOTOMETRY}\\
\hline
\hline
Source &
Wavelength &
Photometry \\
\hline
\hline
Cen A         & 8.8~\micron & 0.9 ($\pm$ 0.1) Jy \\
Cen A         & 10.3~\micron & 1.4 ($\pm$ 0.5) Jy \\
\hline
$\gamma$ Crux & 8.8~\micron & 1009 ($\pm$ 50) Jy \\
IRC$+$10220   & 10.3~\micron & 1.5 ($\pm$ 0.4) Jy \\
\hline
\hline
\end{tabular}
\end{center}
\end{scriptsize}
\end{table}

\clearpage

\figcaption[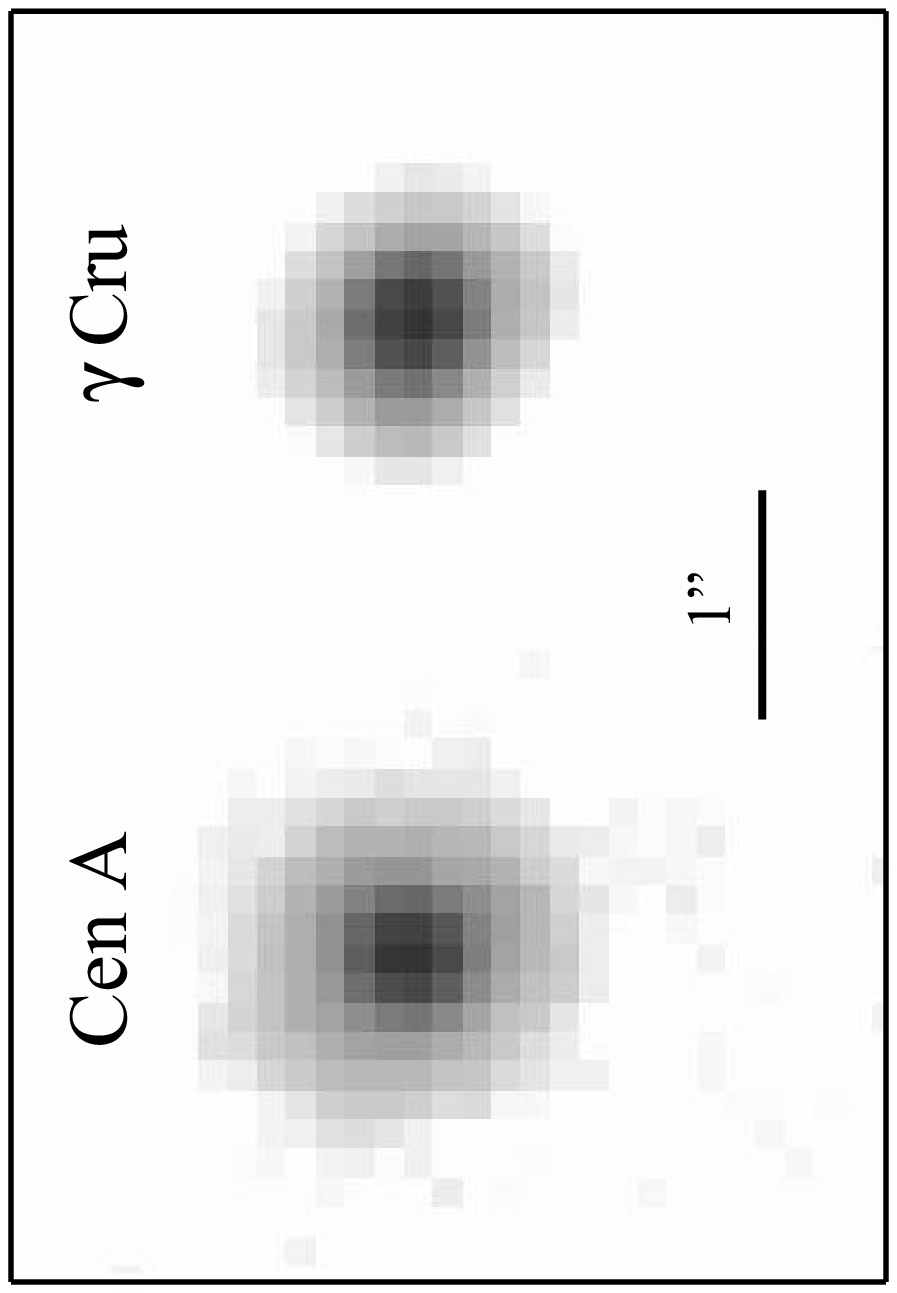]{MIRAC/BLINC Images of Cen A (left) and the reference star
$\gamma$~Crux (right) at 8.8~$\mu$m. North is rotated 126 degrees to
the left (from the vertical)\label{fig1}}

\figcaption[f2.eps]{Radial profiles of Cen A image (solid line) and of the
reference image (dashed line) at 8.8~$\mu$m. The radial
profile of Cen A appears more extended when compared to the radial
profile of the reference star, indicating therefore that Cen A is
resolved.\label{fig2}}

\figcaption[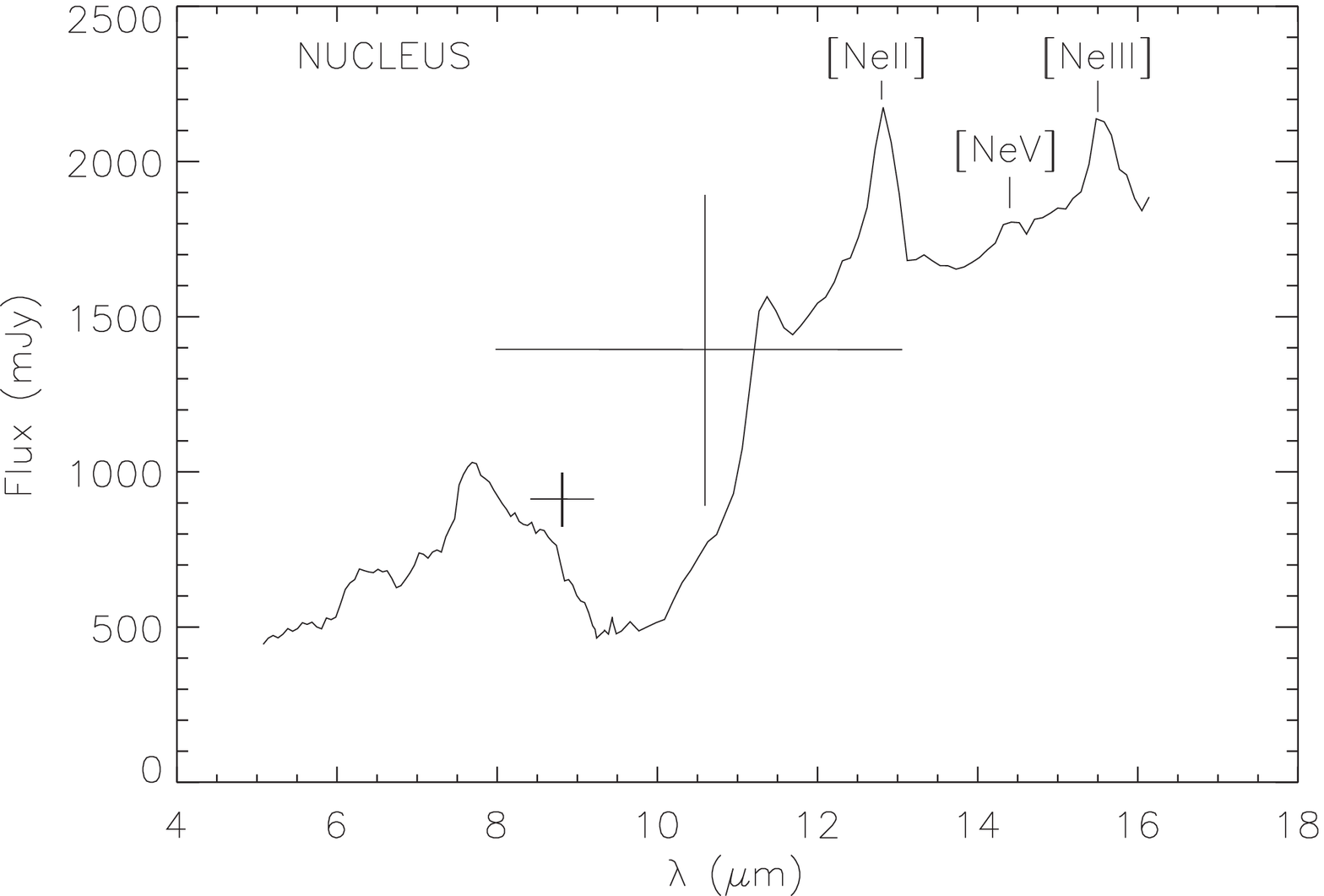]{ISOCAM CVF spectrum of the nuclear region (4"
radius) of Cen A (Mirabel {\it at al.} 1999 showing several emission
lines and a deep silicate absorption feature
at $\sim $9~$\mu$m. Also plotted are the results from our 
8.8$\mu$m and N band photometry of the 2.4'' region centered on the nucleus.\label{fig3}}

\clearpage


\epsscale{0.85} \plotone{f1.eps}  \clearpage
\epsscale{0.85} \plotone{f2.eps}  \clearpage
\epsscale{0.85} \plotone{f3.eps}  \clearpage

\end{document}